\documentclass[conference]{IEEEtran}
\usepackage{times,amsmath,amsthm,color,amssymb,graphicx,epsfig,algorithm,algorithmic,bm,multirow,cite}
\usepackage{lipsum}

\theoremstyle{definition}

\floatname{algorithm}{Algorithm}

\title{Study on Energy Consumption and Coverage of Hierarchical Cooperation of Small Cell Base Stations in Heterogeneous Networks}
\newcommand\blfootnote[1]{%
\begingroup 
\renewcommand\thefootnote{}\footnote{#1}%
\addtocounter{footnote}{-1}%
\endgroup 
}

\author{Xuming Yao$^{\dagger}$, Yingying Sun$^{\dagger}$ and Tao Han$^{\dagger}$\\
$^{\dagger}$ School of Electronic Information and Communications, Huazhong University of Science and Technology,  China\\
Emial: yaoxuming@hust.edu.cn, sunny\_hbqq@hust.edu.cn, hantao@hust.edu.cn}

\IEEEoverridecommandlockouts
\begin{document}
\maketitle

\begin{abstract}
The demand for communication services in the era of intelligent terminals is unprecedented and huge. To meet such development, modern wireless communications must provide higher quality services with higher energy efficiency in terms of system capacity and quality of service (QoS), which could be achieved by the high-speed data rate, the wider coverage and the higher band utilization. In this paper, we propose a way to offload users from a macro base station(MBS) with a hierarchical distribution of small cell base stations(SBS). The connection probability is the key indicator of the implementation of the unload operation. Furthermore, we measure the service performance of the system by finding the conditional probability-coverage probability with the certain SNR threshold as the condition, that is, the probability of obtaining the minimum communication quality when the different base stations are connected to the user. Then, user-centered total energy consumption of the system is respectively obtained when the macro base station(MBS) and the small cell base stations(SBS) serve each of the users. The simulation results show that the hierarchical SBS cooperation in heterogeneous networks can provide a higher system total coverage probability for the system with a lower overall system energy consumption than MBS.
\end{abstract}


\blfootnote{The corresponding author is Yingying Sun. The authors would like to acknowledge the support of the International Science and Technology Cooperation Program of China (Grant No. 2015DFG12580), the National Natural Science Foundation of China (Grant No. 61471180), and the EU FP7-PEOPLE-IRSES (Grant No. 610524). This research is supported in part by the China International Joint Research Center of Green Communications and Networking (No. 2015B01008).}

\section{INTRODUCTION}\label{sec:Intro}
The development of communication technology is no longer just to meet the basic information exchange. In contrast, the richer intelligent terminals and communication models need to have higher speed data rate, more extensive coverage area and higher band utilization as the basic guarantee \cite{Katiyar2011-p409-417}.  A Software - Defined Network (SDN) \cite{Nunes2014-p1617-1634,Kreutz2015-p14-76} based on control plane and data plane separation has been applied in wired communication.  A small cell base station (SBS) with a smaller coverage but less energy consumption than a macro base station (MBS) is hierarchically introduced to a conventional radio base station area where only a macro base station is \cite{Andrews2012-p497-508}.

This article works on the hierarchical distribution of small cell base stations (SBS) offloading users from macro base stations (MBS) cooperatively.  For the hierarchical distribution of small cell base stations, in the same layer of the base station, power consumption should be the same as well, which is different among different layers of SBS, however.  Then all layers of SSB together with the macro base stations forming a heterogeneous network, having been very mature to apply to wireless communication systems \cite{Park2014-p1140-1145}.

In order to improve the performance of the communication system, the introduction of MIMO (Multiple Input Multiple Output) technology is designed to use multiple transmit and receive antennas at the transmitting and receiving ends respectively, so that the signal is transmitted and received through a plurality of antennas at two different ends, thereby the communication quality changing.  However, due to the restrictions in terms of space and spectral range, increasing the number of transmit and receive antennas blindly is impossible \cite{Ding2013-p-}. Besides the MIMO, the converged network is popular to increase the communication system quality as well. In this paper, we propose to improve the density of SBS distribution hierarchically as well as the number of its layer, avoiding the difficulties in achieving high density antennas, which is known as the cooperative communication. Cooperative communication is an obtained example of the converged communication technology.Mobile converged network described in reference\cite{Han201734-p-40} demonstrates the way to converge the mobile networks upon new framework, and the design principles as well as the advantages of the framework are discussed too. The simulation result shows that the normalized average interference increases with increasing normalized intensity of user terminals. 

In this paper, our main contributions are as follows:
\begin{itemize}
\item We propose a hierarchically cooperative communication model where users could be offloaded from MBS to SBS;
\item Analytical results on connection probability, coverage probability and power consumption are obtained, for both MBS and SBS connection;
\item The analysis gives the range of some indicators for better system quality.
\end{itemize}

The rest of this paper is organized as follows: Section II reviews related work. Section III describes the system model. The analysis on the connection probability, coverage probability and power consumption are presented in Section IV. Section V validates the analysis upon simulations. Finally, Section VI concludes this paper.

\section{RELATED WORK}\label{sec:Related Work}
Recently, there are plenty of investigations carried out on cooperative communications.  Reference \cite{Katiyar2011-p409-417} presents a review of cooperative communications.  As is described in the reference, cooperative relaying technique spatially distributes communication nodes, leading to a diversity gain which can be translated into robustness against fading for same transmit power.  Such technique has great potential for high data rate, spectrally efficient and reliable wireless communications. Reference \cite{Peng2011-p-} presents a hierarchical cooperative relay-based heterogeneous network (HCR-HeNet)  based on the technique. The HCR-HeNet divides its coverage into three corresponding layers and is deployed to provide a cost-effective coverage extension based on the convergence of heterogeneous radio networks. The heterogeneous network has been applied in practice. Ubiquitous information service converged by different types of heterogeneous network is one of the fundamental functions for smart cities. Reference \cite{Han2017-p44-50} considers that 5G converged cell-less communication networks are proposed to support mobile terminals in smart cities based on the development of 5G ultra-dense wireless network. In Reference \cite{Ahmed2012-p640-652}, the author denotes a different kind of cooperation communication architecture called smart grid architecture.  A limited number of smart layers are deployed to improve the performance of the communication network and better delivery latency, throughput and energy efficiency are shown as a result.  Reference \cite{Zhou2008-p-} and \cite{Zhou2008-p3618-3628} put more focus on energy-efficiency and compared the efficiency between no-cooperative and cooperative communication.  The resulting performance on energy saving is significant for cooperative communication. The result shows that better cooperative communication schema can reduce the overall energy consumption.  \cite{Han2016-p7381-7392} proposes a method which use small cell cooperation in heterogeneous network to offload users from MBSs. By using software-defined-network (SDN) controller, all the cell association and cooperation are conducted.  Under this circumstance, numerical results show that small cell cooperation can offload more users from MBS tier together with better performance on system's coverage. Differently, this paper puts more focus on the energy consumption and coverage of hierarchical cooperative of small cell base stations in heterogeneous networks and a simulation was made to demonstrate the results.As we have talked, the cell-less communication and the cooperative communication are both converged network technology. In the reference \cite{Wang2017-p-}, the researchers demonstrate the 5G vehicular network upon the cell-less communication. In this paper,  researchers design that moving access points are deployed on vehicles to facilitate the access of  vehicle users.

\begin{figure}[t]
    \centering\includegraphics[width=3.4in]{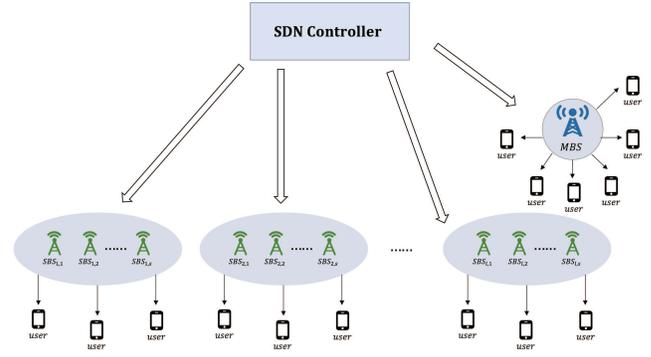}
    \caption{A software defined heterogeneous network with hierarchical small cell cooperation.}\label{fig:systemmodel}
\end{figure}

\section{SYSTEM MODEL}\label{sec:system model}
In this paper, we consider a Software-Defined (SDN) Heterogeneous networks (HetNet) where macro base station (MBS) and hierarchical small cell base station (SBS) coexist, with the different transmission power $p_\mathrm{m}$ and $p_\mathrm{s}$. Two stations both meet the independent 2-D Poisson Point Process (PPP), and the probability density is denoted as $\lambda_\mathrm{m}$ and $\lambda_\mathrm{s}$, respectively. The so-called hierarchical small cell base stations refer to that the transmission power between the small cell base stations from different layers are different. But within a communication area, where are $l$ layers of small cell base stations, the i-th level small cell base stations are the same transmission power,  $p_\mathrm{s,i}$.  The distribution of SBS of each layer satisfies the density of $\lambda_\mathrm{s}$ independent 2-D Poisson process. As we have mentioned the aim of introduction of the SBS is to offload users from the MBS which will take over the communication pressure for the MBS. The connection between the users and the BSs is based on the rule of the maximum received-signal-strength (RSS).  In order to offload users from the MBS more efficiently, we introduce the concept of cell bias, which uses a corresponding scale efficiency, $\beta_\mathrm{i}$, for the SBS with different transmit power so that the result of $\beta_\mathrm{i}p_\mathrm{s,i} (\beta_\mathrm{i}>1)$, larger than the original power, would compete with the RSS from MBS. This means that compared with the MBS, the cooperative SBSs have a greater probability of offloading users, not only for the station number but also for thanks to the bias indicator. For the SBS itself, it is also successful to make sure that the SBSs in the different layers, with different launch power, possibility to offload users in terms of power is not too small, avoiding that the small launch power layer is likely to be idle and causing the waste of resources. System schematic diagram shown in Fig.\ref{fig:systemmodel}.

In the specific transmission process, the BSS uses the orthogonal frequency division multiple (OFDM), the technique eliminating the interference between different BSs (including the interference between different layers of SBS).  We denote $x_\mathrm{m}$ as the location of the MBS and $r_\mathrm{m}$ as the distance between the user and the MBS. $x_\mathrm{s_{i,j}}$ represents the location of the j-th SBS in the i-th layer, $r_\mathrm{s_{i,j}}$ represents the distance between the user and the corresponding SBS. In this paper, we consider a Rayleigh fading channel represented by $h$, which is assumed with zero mean and variance one, i.e., $s(n) \sim \mathcal{CN}(0,1)$. $\alpha>2$ is the path loss exponent for every layer. $h_\mathrm{m}$ represents the channel parameter between the user and the MBS, and $h_\mathrm{s_{i,j}}$ represents the channel parameter between the user and the corresponding SBS.

\section{system indicators}\label{sec:system indicators}
In this section, we will mainly analyze three system indicators, which are connection probability, coverage probability and the energy consumption.
\subsection{Connection Probability}
To get the connection probability, we first define the cooperative cluster $\phi_\mathrm{i}$ which consists of $k$ closest SBS for the layer of $i$. Now we take the i-th layer of SBS as an example to analysis the connection probability. The RSS of a user gets from MBS and SBS cluster from i-th layer are:
\begin{align}
RSS_\mathrm{M}&=p_\mathrm{m}r_\mathrm{m}^{-\alpha}, \label{eq:RSS_m}\\
RSS_\mathrm{s_i}&=\sum_{x_{\mathrm{s_{i,j}\in\phi_\mathrm{i}}}}\beta_\mathrm{i} p_\mathrm{s,i} r_\mathrm{s_{i,j}}^{-\alpha} \nonumber \\
&=k p_\mathrm{s}\sum_{x_{\mathrm{s_{i,j}\in\phi_\mathrm{i}}}}r_\mathrm{s_{i,j}}^{-\alpha}.\label{eq:RSS_s}
\end{align}

For a user, only the RSS of MBS is larger than all SBS clusters’ then it could connect to the MBS. We have denoted that the distribution among the different layers of SBS is independent, which means the probability of MBS’s RSS lager than all SBS clusters’ RSS equals to multiple all probability of MBS lager than every cluster in different layer.  Then we have:
\begin{equation}
P_{MBS}=\prod^{\phi_\mathrm{l}}_{\phi_1}\left(1-P_{SBS_i}\right). \label{eq:P_mbs}
\end{equation}

And from the \cite{Han2016-p7381-7392}, we know the connection probability of a single SBS cluster is:
\begin{align}
P_{SBS_\mathrm{i}} &= \mathrm{Prob} \left\{ \sum_{x_{\mathrm{s_{i,j}\in\phi_\mathrm{i}}}}k p_\mathrm{s}r_\mathrm{s_{i,j}}^{-\alpha} > p_\mathrm{m}r_\mathrm{m}^{-\alpha} \right\} \nonumber \\
&= \int^{+\infty}_{0<r_\mathrm{s_{i,1}}<r_\mathrm{s_{i,2}}<\dots<r_\mathrm{s_{i,k}}}\mathrm{e}^{-\lambda_\mathrm{m}\pi\eta^{2/\alpha}}f_\Gamma(r)dr,\label{eq:P_sbsi}
\end{align}
where $\eta = \frac{p_\mathrm{m}}{\beta_\mathrm{i}p_\mathrm{s}\sum^{x_\mathrm{s_{i,k}}}_{x_\mathrm{s_{i,j}} \in \phi_{\mathrm{i}}} r^{-\alpha}_{\mathrm{s_{i,j}}}} $ and $f_\Gamma(r)$ is the joint probability density function (PDF) of $r_\mathrm{s_{i,j}}$. It is given by:
\begin{equation}\label{eq:f_pdf}
f_\Gamma(r)=\left(2\pi\lambda_\mathrm{s}\right)^{k}\mathrm{e}^{-\lambda_\mathrm{m}\pi r^{2}_\mathrm{s_{i,j}}} \prod^{x_\mathrm{s_{i,k}}}_{x_\mathrm{s_{i,j}}\in\phi_\mathrm{i}}r_\mathrm{s_{i,j}}.
\end{equation}
Thus, from \eqref{eq:P_mbs} and \eqref{eq:P_sbsi}, we have:
\begin{equation}\label{eq:P_mbs_expand}
P_{MBS}= \prod^{\phi_\mathrm{l}}_{\phi_1}\left(1-\int^{+\infty}_{0<r_\mathrm{s_{i,1}}<r_\mathrm{s_{i,2}}<\dots<r_\mathrm{s_{i,k}}}\mathrm{e}^{-\lambda_\mathrm{m}\pi\eta^{2/\alpha}}f_\Gamma(r)dr\right).
\end{equation}

For the SBS as a whole, we can get the connection probability of the system as:
\begin{equation}\label{eq:P_sbs}
P_{SBS}=1-P_{MBS}.
\end{equation}

Considering the cooperative contribution from SBS clusters, it is very reasonable that the RSS of SBS clusters are easily larger than MBS, especially we also use the cell bias scaler $\beta_\mathrm{i}$. Hierarchical cooperation of SBS could improve the connection probability significantly as well as the coverage probability mathematically. Because a the basement of SBS cooperatively offloading users from MBS is lager distribution density for SBS. From this aspect, the system could provide better connection for marginal users, which is the determined factors for the coverage area. But in other words, the cooperation between the SBS means more information transform among the system, in other word, more complex transform  procedure will result in new energy consumption as well as more complicated system construction.

\subsection{Coverage Probability}
Coverage probability could be understood in this way: after the user connecting to a MBS or a SBS cluster, the conditional probability of the signal received by the user is bigger than a threshold $\theta$, which is the smallest SINR for a successful communication procedure. In other word, only if the SNR is larger than the threshold $\theta$,  a successful communication link has been constructed.

We have pointed out that the interference between the BSs only within the layer thanks to OFDM system. So, it is not difficult to have the received signal expression:
\begin{equation}
\sum_{x_\mathrm{i,j}\in\mathcal{B}}\frac{ \left( Ph_\mathrm{i,j} \right)^{1/2} }{r^{\alpha/2}_\mathrm{i,j}}X+ \sum_{x_\mathrm{i,j}\notin\mathcal{B}}\frac{ \left( Ph_\mathrm{i,j} \right)^{1/2} }{r^{\alpha/2}_\mathrm{i,j}}Y+Z,
\end{equation}
where $P$ is the transmit power of the serving BS, this parameter of the MBS and SBS are not the same, and the former one is much larger than the latter one; $\mathcal{B}$ represents the set of BSs to which the user is connected and can be a single MBS or SBS cluster; $X$ represents the input symbol sent by the connected BS, that is, the useful signal. Thus, the first term of the formula represents the useful signal from the associated BS, and the summation will degrade to a single value when the MBS is connected. $Y$ denotes the input symbol sent by the BSs that do not belong to $\mathcal{B}$ ; $Z$ is a circular symmetric zero mean complex Gaussian random variable with variance $\sigma^2$. Because it is additive white noise and the additive white noise represents the influence of weather, temperature or other forms of interference throughout the environment. From the definition of the SINR, we have the SINR of the system as:
\begin{equation}\label{eq:SNR}
\mathrm{SNR}(\mathcal{B})=\frac{\left|\sum_{x_\mathrm{i,j}\in\mathcal{B}}(Ph)^{1/2}r^{-\alpha/2}\right|^2}{\left|\sum_{x_\mathrm{i,j}\notin\mathcal{B}}(Ph)^{1/2}r^{-\alpha/2}\right|^2+\sigma^2}.
\end{equation}

Thus, for a given SNR threshold, the definition of the coverage probability at a typical user can be easily expressed as:
\begin{equation}\label{eq:coverage_prob}
\mathcal{P}=\mathrm{Prob}\{\mathrm{SNR}>\theta\}.
\end{equation}

Based on the reference [7], we have the coverage probability for a single SBS cluster as \eqref{eq:coverage_i}
\begin{figure*}[t]
\begin{equation}\label{eq:coverage_i}
\mathcal{P}_\mathrm{s}(i) = \int^{+\infty}_{0<r_\mathrm{s_{i,1}}<r_\mathrm{s_{i,2}}<\dots<r_\mathrm{s_{i,k}}}
 \exp \left[ { \left( \frac{-\theta\sigma^2}{\sum^{x_\mathrm{s_{i,k}}}_{x_\mathrm{s_{i,j}}}} \right)} {\mathcal{L}_\mathrm{I}\left( \frac{\theta}{\sum^{x_\mathrm{s_{i,k}}}_{x_\mathrm{s_{i,j}}}}\right)}\right]f_\mathrm{R_cs}(r)dr.
\end{equation}
\hrulefill
\end{figure*}
where $f_\mathrm{R_{cs}}=\frac{1}{P_{SBS}}\mathrm{e}^{-\pi\lambda_\mathrm{m}\eta^{2/\alpha}}f_{\Gamma}(r)$ is the PDF of the distance between the user and the SBS. $\mathcal{L}_\mathrm{I}(s)$ is the distance distribution [7].

Then, the whole coverage probability for SBS goes as
\begin{equation}\label{eq:coverage_whole}
\mathcal{P}_\mathrm{s}=\sum^{\phi_\mathrm{m}}_{\phi_1}\mathcal{P}_\mathrm{s}(i).
\end{equation}

Similarly, for the MBS we have:
\begin{align}
\mathcal{P}_\mathrm{m}(m) &= \int^{+\infty}_{0}\left[\mathrm{e}^{-P^{-1}_\mathrm{m}\theta^{\alpha}_{r}}\right]f_{\mathrm{R_cm}}(r)dr \label{eq:prob_m}\\
f_{\mathrm{R_cm}}(r) &= \frac{f_{\mathrm{r_m}}(r)\mathrm{Prob}\left( \frac{P_\mathrm{m}r^{-\alpha}}{P_\mathrm{s}}>\sum^{x_\mathrm{s_{s,k}}}_{x_\mathrm{s_{s,j}}\in\mathcal{C}}r^{-\alpha}_{\mathrm{s,j}}\right)}{P_{MBS}}.\label{eq:pdf_rcm}
\end{align}

Therefore, along with the connection probability, it is easy to derive the user's total coverage probability of the model:
\begin{equation}\label{eq:total_coverage}
\mathcal{P}=P_{MBS}\mathcal{P}(m)+P_{SBS}\mathcal{P}_\mathrm{}.
\end{equation}

In this case, it is not difficult to analyze the expression of SINR and the probability density expression of the distance between the user and the BS. If the user is associated with only one BS, its second nearest neighbor BS becomes its strongest interference naturally, and the cooperative communication just makes good use of this interference. So it is easy to accept that: with the increase in the size of the cluster, this advantage will be more obvious, but at the same time will increase the cost of the system to improve communication coverage and quality.

\subsection{Energy Consumption}
In the communication system, the energy cost is never to be overlooked as a major aspect, which is not only the basic requirements of green communication, but also is closely related to the main part in terms of the production activities and economic benefits. To this end, in order to study the energy consumption of the system better, the specific energy consumption is divided into two parts as the system self-cost and service overhead. Basing on this, we could get the expression of the energy consumption of MBS and a SBS cluster respectively:
\begin{align}
P_{mbs} &= P_{ms} + \frac{n P_{m}p_{MBS}}{N},\label{eq:energy_Pmbs}\\
P_{sbs_\mathrm{i}} &= knp_{SBS}(P_\mathrm{s}+P_\mathrm{bkh}). \label{eq:energy_Psbs}
\end{align}
So, for the SBS as a whole we have:
\begin{equation}\label{eq:energy_total}
P_{sbs}=\sum^{\phi_\mathrm{l}}_{\phi_1}knp_{SBS}(P_s+P_{bkh}).
\end{equation}

$P_{ms}$ and $P_\mathrm{s}$ are static power consumption, mainly for some basic circuit units, such as processing units and radio modules; $n$ is the number of users within the region; $N$ is in the protection of system quality of service under the premise of MBS is fully loaded when the maximum number of users; $P_{bkh}$ is the return of each cooperative SBS power consumption, it is this part of the energy costs from interference into a useful signal, the advantages of collaborative communication is so obvious.

\section{SIMULATION}

\begin{figure}[t]
    \centering\includegraphics[width=2.8in]{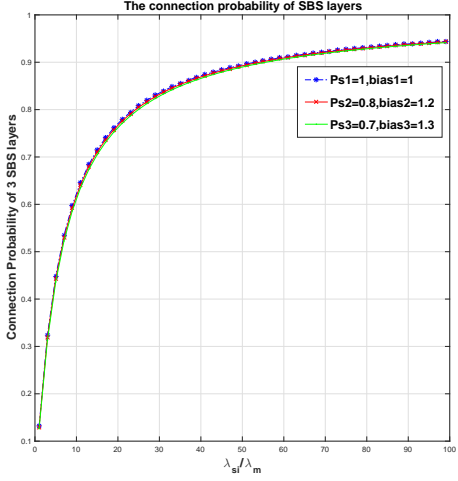}
    \caption{The connection probability of SBS layers.}\label{fig:result1}
\end{figure}

In this section, we will present numerical evaluation of the results presented in the previous section, then make some discussion and conclusions upon our simulation results. Without losing generality, we assume there are 10 different SBS layers and set the number of the SBS in a cluster to be 2 to analysis the results. First, we see the Fig. \ref{fig:result1}.

We analyze the SBS layers' connection probability with different $\beta_\mathrm{i}$, $\lambda_\mathrm{s}$ and$\lambda_\mathrm{m}=(500^2\pi)^{-1}\theta=5$. As we mentioned above, here 2 cooperative SBSs in one cluster are considered. During our MATLAB simulation, we use a 2-D integral to simplify the range of the integration.

From the result, we could clearly notice with the growing of $\lambda_\mathrm{s}$ , SBS's distribution density , the connection probability for different layers increase similarly. To offload more users from the MBS, not only could we enlarge the  size of cooperative cluster but also increase the distribution density of the SBS for different layers. The result illustrates that when $\lambda_\mathrm{s}$/$\lambda_\mathrm{m}$ is around 25, it is obviously for a user to connection to SBS cluster whose connection probability is lager than 0.8.

\begin{figure}[t]
    \centering\includegraphics[width=2.8in]{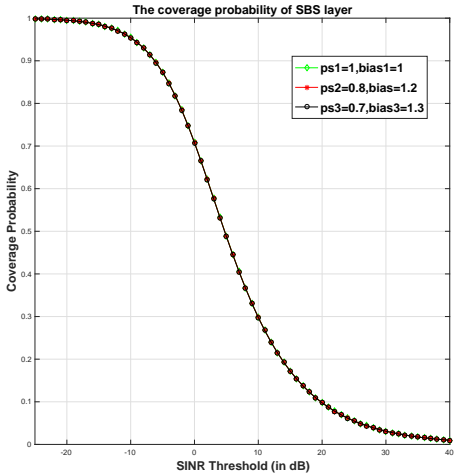}
    \caption{The coverage probability of SBS layers.}\label{fig:result2}
\end{figure}
Then, we move to  the Fig. \ref{fig:result2},in which the SBS layers' coverage probability with different $\beta_\mathrm{i}$, $\lambda_\mathrm{s}$ =50$\lambda_\mathrm{m}$and$\lambda_\mathrm{m}=(500^2\pi)^{-1}$.In this result, the coverage probabilities for SBS clusters from different layers decrease with SNR threshold increasing. In other word, if the lowest communication power threshold increases,which can provide better communication stability, the coverage influence will be weaken naturally.
\begin{figure}[t]
    \centering\includegraphics[width=2.8in]{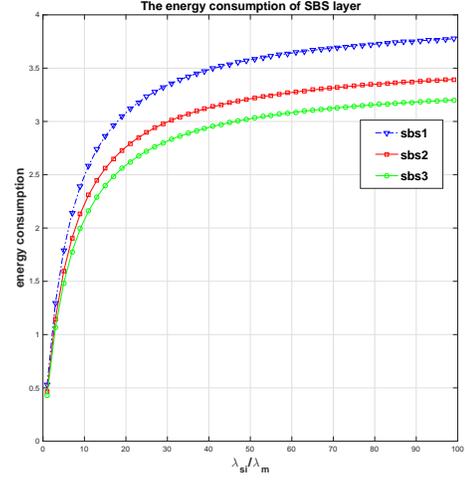}
    \caption{The energy consumption of SBS layers.} \label{fig:result3}
\end{figure}

From the simulation results in Fig. \ref{fig:result3}, we analyze the energy consumption in the Fig. \ref{fig:result1}. It is reasonable that the power consumption and $\lambda_\mathrm{s}$ are positive correlated. From the result, we notice when  correlation between two indicators becomes less when the $\lambda_\mathrm{s}$/$\lambda_\mathrm{m}$ is larger than  50, which is constrained by the size of the cluster. 

\begin{figure}[t]
    \centering\includegraphics[width=2.8in]{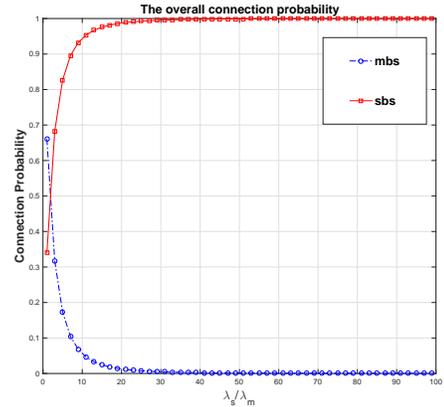}
    \caption{The overall connection probability.}\label{fig:result4}
\end{figure}

Above 3 simulation results only come from the bias SBS cooperative clusters, in following part, we will focus on comparing the indicators between SBS and MBS and the first is the connection probability, which is shown in Fig. \ref{fig:result4}.

In the Fig. \ref{fig:result4}, we calculate the overall connection probability  with different $\lambda_\mathrm{s}$/$\lambda_\mathrm{m}$ and $\lambda_\mathrm{m}=(500^2\pi)^{-1} ,\theta=5$.In a Heterogeneous Networks, for a user, it could only connect to the SBS cluster or the MBS, so the sum of those two connection probability naturally equals to 1. And from the simulation results, we can clearly see when the  $\lambda_\mathrm{s}$/$\lambda_\mathrm{m}$ is lager than 20, SBS cluster could offload the users from the MBS with the probability near 1. 

\begin{figure}[t]
    \centering\includegraphics[width=2.5in]{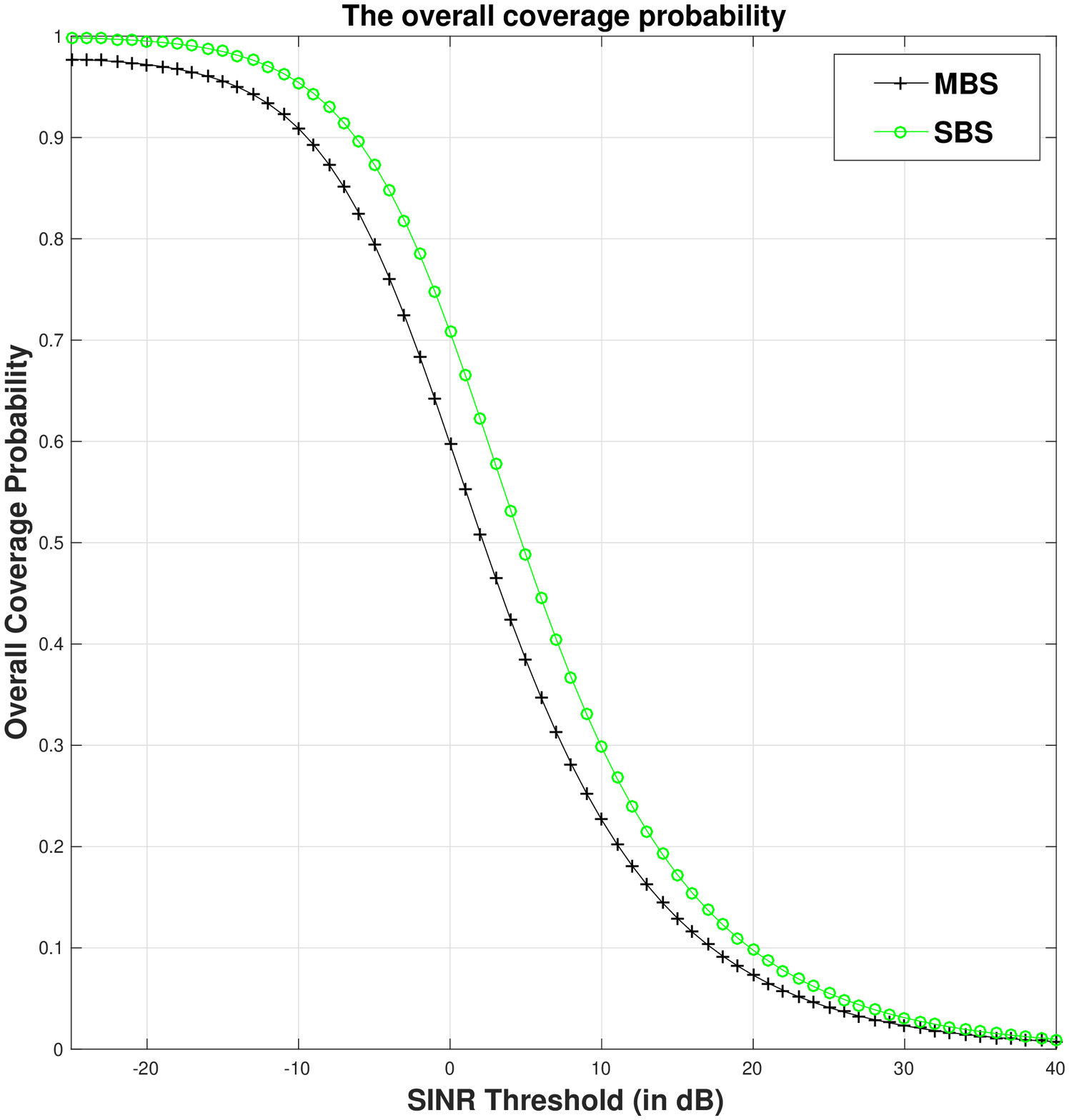}
    \caption{The overall coverage probability.}\label{fig:result5}
\end{figure}

In the Fig. \ref{fig:result5}, we notice the overall coverage probability with increasing the SNR threshold. For two BS, larger SNR threshold will decrease the system coverage area. But it is still obvious  that the SBS cluster will provide better coverage service as green line shows in the  Fig. \ref{fig:result5}.

\begin{figure}[t]
    \centering\includegraphics[width=2.4in]{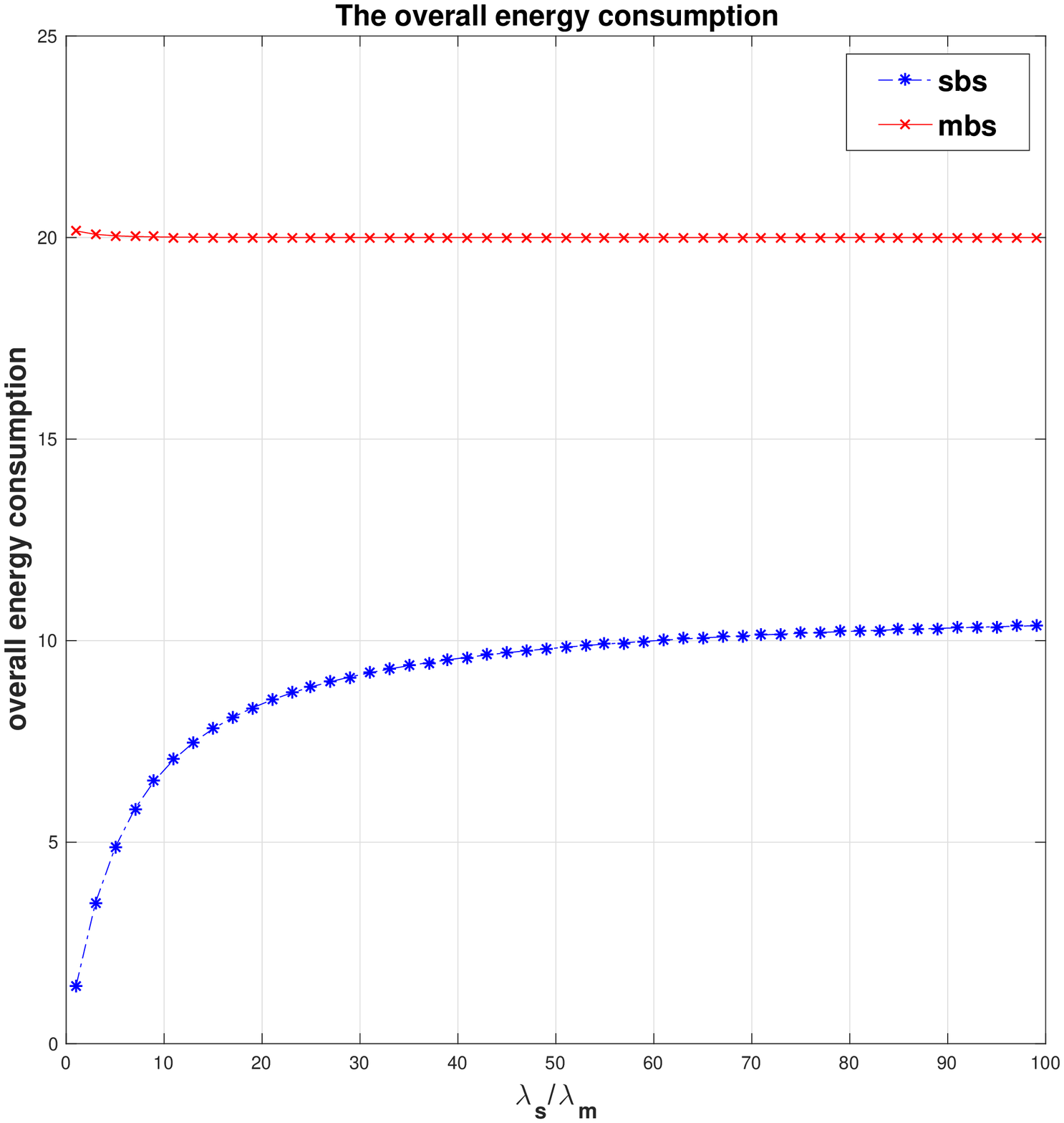}
    \caption{The overall energy consumption.}\label{fig:result6}
\end{figure}

In the Fig. \ref{fig:result6}, the energy consumption of the SBS and MBS differ from each other significantly. When the user is covered by MBS, the  energy consumption is stable  through out the  $\lambda_\mathrm{s}$/$\lambda_\mathrm{m}$ values while the indicator in terms of SBS increases evidently  before the  $\lambda_\mathrm{s}$/$\lambda_\mathrm{m}$  goes to 50. And from the previous results, we can also point out that the bias SBS cluster could provide better connection probability as well as better coverage probability only with the half energy consumption compared with the MBS.

\section{CONCLUSION}\label{sec:conclusion}
Hierarchical distribution of SBS cooperative communication system can greatly improve the system coverage probability and energy efficiency. By introducing the concept of cell bias, the performance of SBS is stronger, and then the system indicators are improved at the same time. Increasing the SBS distribution density is an effective way to solve the heavy load of the MBS. The hierarchical SBS further optimizes the offloading performance with cooperative SBS clusters. The connection between the users and the BS is based on the RSS rules, and the establishment of the user-cantered research model has a great reference significance relative to the actual situation. The joint use of heterogeneous network and software definition network technology is the key to the realization of the entire system. The control plane and data plane separating from each other is a good idea of the development of communication technology.

\bibliographystyle{IEEEtran}
\bibliography{sample}
\begin{IEEEbiography}[{\includegraphics[width=1in,height=1.25in,keepaspectratio]{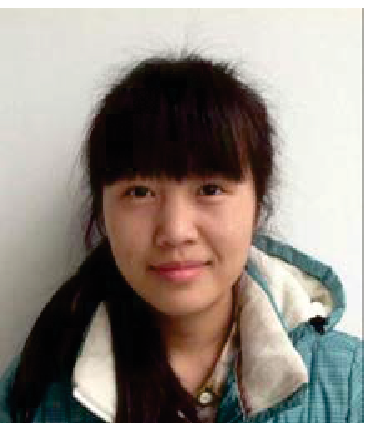}}]{Xuming Yao}
Xuming Yao  received the bachelor’s degree in Electronic and Information Engineering from Huazhong University of Science and Technology(HUST), Wuhan, China, in 2016. She is currently working toward the Master’s degree in the University of Edinburgh, the United Kingdom. Her research interests include cooperative communication and signal processing.
\end{IEEEbiography}

\begin{IEEEbiography}[{\includegraphics[width=1in,height=1.25in,keepaspectratio]{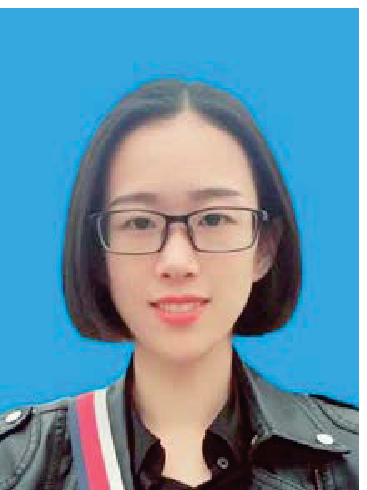}}]{Yingying Sun}
Yingying Sun received the Bachelor’s degree in
electronic and information engineering from Wuhan
University of Technology, Wuhan, China, in 2016.
She is currently working toward the Master’s degree
in Huazhong University of Science and Technol-
ogy (HUST), Wuhan, China.Her research interests
include cooperative communication and vehicular
network.
\end{IEEEbiography}

\begin{IEEEbiography}[{\includegraphics[width=1in,height=1.25in,keepaspectratio]{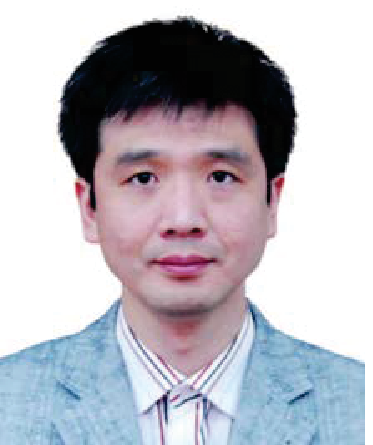}}]{Tao Han}
Tao Han received the Ph.D. degree in informa-
tion and communication engineering from Huazhong
University of Science and Technology (HUST),
Wuhan, China, in 2001.
He is currently an Associate Professor with the
School of Electronic Information and Communica-
tions, HUST. From 2010 to 2011, he was a Visiting
Scholar with the University of Florida, Gainesville,
FL, USA, as a Courtesy Associate Professor. He
has published more than 50 papers in international
conferences and journals. His research interests in-
clude wireless communications, multimedia communications, and computer
networks.
He is currently serving as an Area Editor for the European Alliance
Innovation Endorsed Transactions on Cognitive Communications.
\end{IEEEbiography}

\end{document}